\documentclass[11pt]{article}
\usepackage{moriond,epsfig}

\def\Journal#1#2#3#4{{#1} {\bf #2}, #3 (#4)}

\def\NIMA{{\em Nucl. Instrum. Methods} A}

\def\be{\begin{equation}}
\def\ee{\end{equation}}
\def\bea{\begin{eqnarray}}
\def\eea{\end{eqnarray}}

\begin{document}
\vspace*{4cm}
\title{STATUS OF MAGIC AND EXPECTED CAPABILITIES}

\author{J. CORTINA for THE MAGIC COLLABORATION}

\address{Max-Planck-Institut f\"ur Physik, F\"ohringer Ring 6, 
Munich, Germany\\E-mail: cortina@mppmu.mpg.de}

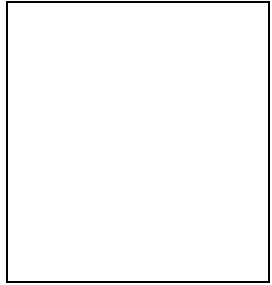
\maketitle\abstracts{
An overview of the status of the 17 m diameter 
MAGIC telescope project will be given. During 
phase I, the telescope will reach a threshold 
of 30 GeV and a sensitivity of 6.0 10$^{-11}$ 
cm$^{-2}$ s$^{-1}$. First light is foreseen in mid 2001
and first observations in 2002. The expected
capabilities of the telescope for high energy
astrophysics and fundamental physics will be 
reviewed.}

\section{Introduction}

Imaging Air Cherenkov Telescopes (IACTs)
have established in recent years 
the existence of a few galactic
and extragalactic sources above 300 GeV~\cite{smith}
(the so-called Very High Energy gamma ray range).
New technical developments currently allow to reduce
the threshold of this kind of telescopes
to energies around 30 GeV, thus overlapping 
those typical of $\gamma$-ray satellites like 
EGRET or the forthcoming GLAST and AGILE.
Here we describe a new generation 17 m diameter
IACT dubbed MAGIC~\cite{lorenz,proposal} (Major Atmospheric Gamma Imaging
Telescope) with a threshold of 30 GeV in its
first development phase (phase I) and around 10 GeV
in phase II.

\section{The general design}

MAGIC is a second generation IACT with a number of improvements 
and innovative elements to reduce the energy threshold and
the different backgrounds. 

\subsection{The reflector}

Thanks to its 17 m diameter disk, MAGIC will collect 
three times more light than a typical first generation 10 m diameter IACT.
The shape of the reflector is parabolic, hence isochronous: this 
improves the light background reduction.
A carbon fiber mount, weighing less than 10 tons,
is instrumental in rapid repositioning the telescope in the search for
GRBs. This material makes the disk also hard to deform (sagging $<$ 3mm).
In addition, an active mirror control system~\cite{active}
provides maximum shape stability.
The reflector is tessellated, with 50 cm $\times$ 50 cm 
all-alluminium, diamond-turned, quartz-coated,
mirror elements~\cite{mirrors}
of 85\% reflectivity in the 300-650 nm wavelength region.

\subsection{The camera}

New compact (2.5 cm, 0.1$^{\circ}$ diameter) photomultipliers~\cite{pmt}
with approximately 20\% quantum efficiency (QE) in the 300-500 nm range
and minimal time dispersion shall be used in phase I. In order to
minimize the impact of DC like night sky background (NSB) 
and moonlight the dynodes
were reduced to six and operated with a total gain of only 15000.
The camera of 4$^{\circ}$ diameter is composed of 600 PMTs
with coarser sampling in the outer region.

In phase II the camera will be provided with Hybrid Photo Detectors
of $<$QE$>$ $\sim$ 40\% 
in an wider wavelength range of 330-650 nm. Extending
the light detection range to red wavelengths is particularly
useful for high zenith angle observations where most
of the short wavelength photons are absorbed.
Finally, in its phase III, MAGIC aims at equipping the
camera with Avalanche Photo Diodes, provided with an 
extremely high $<$QE$>$ $\sim$ 80\% (330-680 nm).

\subsection{The readout}

Cherenkov light pulses are typically very short 
(in the order of 1-2 ns).
However as the pulse shape contains important physical information 
and can be used for NSB and hadron reduction, it is
crucial to register it faithfully.
MAGIC endeavors to do so by applying a number of
improvements. To begin with the signal is transported 
over optical fiber~\cite{analog}. Apart from negligibly dispersing the
pulse, optical transport reduces cable weight and enables
optical decoupling and noise immunity.

In addition, the pulse is digitized using very fast (300 MHz,
and 1 GHz in a second phase) 
8 bit FlashADCs~\cite{fadc}. The maximum sustained event
rate will be 1 kHz (with zero deadtime). Dual ranging  
of the analog signals at the FADC inputs extends the
range of the FADCs to about 70-80 dB.

\subsection{Sensitivity and performance of MAGIC phase I}

The telescope will have a threshold of about 30 GeV
(peak of the differential flux) for phase I.
The gamma-ray sensitivity ranges from 
$\sim$10$^{-10}$ at around 10 GeV 
(set by the cosmic electron background~\cite{electrons})
down to 8 $\cdot$ 10$^{-12}$ cm$^{-2}$ s$^{-1}$ at 1 TeV
(determined by the cosmic hadron background).
The collection area flattens at around 10$^5$ m$^2$ above
100 GeV. Close to the threshold the energy resolution
will be around 30\% and improve 
to around 10\% at 1 TeV~\cite{energy}.  

In the presence of moonlight, the threshold has to
be increased to 60-100 GeV (half-moon, $>$ 30$^{\circ}$ away).
Operation under moonlight significantly increases 
the potential of the telescope to detect GRBs
and follow up other transient sources.

MAGIC is under construction in the Roque de los Muchachos
Observatory at La Palma (Canary Islands, Spain). 
First light is expected for summer 2001 and 
first physics results may be available on 2002.

\section{Prospects for Astrophysics and Fundamental Physics}

Here are briefly described some of the 
astrophysics, cosmology and fundamental physics
questions MAGIC is expected to address.

\subsection{Active Galactic Nuclei}

AGNs exhibit their highest variability in 
X-rays and $\gamma$-rays. Multiwavelength campaigns 
including IACTs have already proved central to understand
the physics of the source, specially because
GeV $\gamma$-rays probably come from very close to 
the supermassive black hole. 

\subsection{The cosmological $\gamma$-ray horizon}
 
High energy $\gamma$-rays are absorbed in the infrared background
($\gamma_{VHE} + \gamma_{IR} \rightarrow e^- + e^+$).
This means that there is a maximum observable redshift, 
z$_{hor}$(E), which is usually referred to as $\gamma$-ray horizon.
Only AGNs below this redshift are observable.
Due to its extremely low energy threshold, MAGIC will see the 
bulk of the cosmological AGNs. A large number of
detections shall allow to characterize the AGN population
as a function of distance and determine if $\gamma$-rays
are indeed only externally absorbed or there is 
some internal absorption.  If a large enough number
of AGNs are detected, a measure of the 
infrared background may also be attempted.

\subsection{The diffuse extragalactic $\gamma$-ray background}
 
EGRET has measured the $\gamma$-ray background up to 100 GeV. 
The general opinion is that it is due to AGNs.
But it has been proved~\cite{chiang} that AGNs only contribute
25\% of the background above 100 MeV.
The remaining background may come from FSRQs and BL Lacs~\cite{muecke}.
In this case MAGIC may detect more than 100 more sources 
than EGRET thanks to its higher sensitivity.
The background could alternatively arise from topological defects 
(which may in turn be also the origin of the cosmic rays 
above 10$^{19}$ eV).

\subsection{Pulsars}
 
Three pulsars are among the strongest EGRET sources. 
Observations at energies above 10 GeV may help clarify
the mechanisms producing the $\gamma$-rays.
The outer gap or polar cap models differ in
the maximum energy attainable for the pulse emission.
In addition, and owing to its improved angular
resolution, MAGIC can clarify if many of the 
$\sim$ 100 EGRET unidentified sources
are associated to pulsars. A special pulsar trigger
has been developed to attain a low
energy threshold around 10 GeV based on pulsar timing analysis.

\subsection{The origin of cosmic rays: shell-type SNRs}
 
It is widely believed that CRs are produced in SNR blast shocks.
X-rays and $\gamma$-rays are essential probes, since,
unlike CRs, they point to the source.
There is good evidence of electron acceleration. 
Synchrotron X-rays and inverse Compton $\gamma_{VHE}$'s 
produced by electrons have been observed in SN-1006.
However no firm evidence has been found for 
hadron acceleration. Observation of 
$\gamma$-rays coming from dense clouds serving
as targets for cosmic rays accelerated in the
shell (through $p,Fe + N \rightarrow \pi's$ 
and $\pi^0 \rightarrow \gamma\gamma$) 
would provide an unambiguous proof of the presence
of hadrons in the shell.

\subsection{Gamma Ray Bursts}
 
The mechanism producing GRBs have been not resolved yet. 
MAGIC counts on its huge collection area to
detect fast transients with high statistics.
The telescope has actually been optimized for 
GRB searches by ways of its fast repositioning capability.
Its potential for GRB detection is illustrated by
the very high rates which would have been expected
for some typical GRBs. For example, GRB-910503, with a
maximum measured energy of 10 GeV and lasting for 84 s,
would have produced a rate of 6$\cdot$10$^{4}$ Hz
in MAGIC, whilst GRB-930131, reaching at least 1.2 GeV
and lasting for 100 s, would have produced 
around 2$\cdot$10$^{4}$ Hz. Such strong signals
allow us to study the details of the
light curve and further characterize the source structure.

\subsection{Invariance of the speed of light}
 
Several quantum gravity models predict an energy dispersion 
of the speed of light~\cite{amelino}. This results in
a arrival time difference of high energy photons as respect to 
low energy ones which depends on particle energy and 
quantum gravity energy scale. By detecting 
sub-second time differences between e.g. keV
and 30-300 GeV photons for z$\sim$0.1-1, 
energy scales in the order of 10$^{18}$-10$^{20}$ GeV
are well measurable. A similar analysis may
also be applied to pulsar pulsations.

\subsection{Cold dark matter}
 
One of the most plausible candidates for dark matter 
is the neutralino ($\chi$).
The particle physics lower limit for m$_\chi$ is around 30-50 GeV
and further evidence suggests that its mass should be below 
1 TeV. An annihilation line ($\chi \chi \rightarrow \gamma \gamma$)
may thus be expected from the center of our Galaxy~\cite{bergstrom}
in the most sensitive range of the telescope.

\section*{Acknowledgments}
I am deeply grateful to my colleagues from the MAGIC 
collaboration for providing extensive information.
In particular I would like to thank N. Magnussen,
E. Lorenz and R. Mirzoyan for their kind help.
The support of the German BMBF and Spanish CICYT
is also acknowledged.

\end{document}